Article

# Scientific impact evaluation and the effect of self-citations: mitigating the bias by discounting h-index

## Emilio Ferrara

Center for Complex Networks and Systems Research, School of Informatics and Computing, Indiana University Bloomington, USA

## Alfonso E. Romero

Department of Computer Science, Centre for Systems and Synthetic Biology, Royal Holloway, University of London, UK

**Abstract**

In this paper, we propose a measure to assess scientific impact that discounts self-citations and does not require any prior knowledge on the their distribution among publications. This index can be applied to both researchers and journals. In particular, we show that it fills the gap of h-index and similar measures that do not take into account the effect of self-citations for authors or journals impact evaluation. The paper provides with two real-world examples: in the former, we evaluate the research impact of the most productive scholars in Computer Science (according to DBLP); in the latter, we revisit the impact of the journals ranked in the "Computer Science Applications" section of SCImago. We observe how self-citations, in many cases, affect the rankings obtained according to different measures (including h-index and ch-index), and show how the proposed measure mitigates this effect.

## 1. Introduction

The h-index is a *de facto* standard for measuring academic performance, although its introduction as a tool to evaluate the impact of researchers raised some concerns in the scientific community. Before discussing these limits, we provide some insights on the rationale behind its functioning. The h-index is defined as follows (Hirsch, 2005):

**Definition 1.** The index h is defined as the number of papers of a given author with a number of citations equal or greater than h.

The h-index has been proposed to address a number of typical disadvantages of single-number evaluation criteria, such as total number of published papers per author, total number of citations collected by an author or by her/his most representative publications, and so forth.
Most of these single-number criteria ignore the significance of published papers, to favor instead lower productivity or seniority of authors, or to randomly favoring (or penalizing) authors, depending on arbitrary threshold parameters (Hirsch, 2005).
Both h-index – and its variants (Tol, 2009), (Alonso, Cabrerizo, Herrera-Viedma & Herrera, 2009), (Egghe, 2010), (Van Noorden, 2010) – and a measure called *citations-per-publication index*, can deal with these issues (Lehmann, Jackson & Lautrup, 2006). On the other hand, both of them suffer from major drawbacks, as highlighted by recent literature (Ball, 2007), (Costas & Bordons, 2007), (Schubert & Glanzel, 2007).

**Corresponding author:**
Emilio Ferrara, School of Informatics and Computing, Indiana University Bloomington, 919 E. 10th St., Bloomington, IN 47408 US
Email address: ferrarae@indiana.edu Telephone number: +1 812 856 7841

Alfonso E. Romero, Department of Computer Science, Centre for Systems and Synthetic Biology, Royal Holloway, University of London, Egham, TW20 0EX, UK
Email address: aeromero@cs.rhul.ac.uk

## Emilio Ferrara and Alfonso E. Romero

The citations-per-publication index is biased towards authors publishing few papers that acquire a lot of citations – which is not necessarily an indication of research impact. For example, think of authors of famous books or survey papers that tend to be highly cited due to the broad audience of readers they address. In addition, when applied to determine the impact of scientific journals, the adoption of this measure penalizes those journals publishing original research works rather than surveys or reviews.

The motivations behind the concerns on the adoption of the h-index are subtler. In fact, some of the above-mentioned drawbacks do not apply to the h-index: it is designed to deal with authors that publish few highly cited papers or with those journals publishing few broad interest works.

Unfortunately, the h-index is not exempt of limits – see, for instance (Costas & Bordons, 2007), (Zhivotovsky & Krutovsky, 2008), (Bartneck & Kokkelmans, 2011). To name a few: (i) it does not take into account the number of authors per publication or their order; (ii) it does not consider the years of activity of each given scholar/journal; (iii) it does not take into consideration the scientific field of each given work/journal; and, finally (iv) it can suffer from bias due to the effect of self-citations.

A vast number of variants of the h-index have been proposed during the latest years, to address the shortcomings of the original measure (Tol, 2009), (Alonso et al., 2009), (Egghe, 2010). For example, regarding point (i), hI-index (Batista, Campiteli & Kinouchi, 2006) and hm-index (Schreiber, 2008) modify the original h-index taking into account the number of authors per paper. As for the shortcoming at point (ii), age-weighted citation indicators such as AWCR, AWCRpA (Jin, 2007) and AW-index have been recently proposed. Concerning point (iii), different papers try to assess the impact of research field and its citation habits on the evaluation of the h-index of authors (Batista, Campiteli & Kinouchi, 2006), (Ball, 2007), (Tol, 2009), (Kulkarni, Aziz, Shams & Busse, 2011) and journals (Schubert & Glanzel, 2007), (Alonso et al., 2009).

The last point, i.e., the effect of self-citations on the computation of h-index represents the main topic of this paper. First of all, let us formally define the concept of self-citation as follows:[1,2]

**Definition 2.** A self-citation $C^S_{P \rightarrow Q}$ is any citation appearing in a paper P pointing to paper Q, whose set of authors are respectively *A[P]* and *A[Q]*, for which it holds true: $A[P] \cap A[Q] \neq \emptyset$, i.e., the intersection of the sets of authors is not empty.

With the availability of bibliographical services containing citation data, the quantitative analysis of influence of self-citations in known indexes has recently acquired a relevant attention (Zhivotovsky & Krutovsky, 2008), (Bartneck & Kokkelmans, 2011). In fact, differently from what stated by Hirsch (2005), it has been recently proven that the effect of self-citations on the h-index is relevant and introduces a bias in the impact evaluation (Fowler & Aksnes, 2007), (Testa, 2008). However, some authors observed that the h-index is less biased in some specific research fields in which self-citations occur at lower rates (Engqvist & Frommen, 2008). Other authors noted the opposite effect, for most of the fields in which self-citations heavily bias the outcome of the impact evaluation (Zhivotovsky & Krutovsky, 2008), (Gianoli & Molina-Montenegro, 2009).

In the literature, different motivations have been proposed to explain the presence of self-citations, for example (i) in trending scientific fields, a tight core of publications tend to be highly cited by those same authors working in the field (Schreiber, 2007a); (ii) in those fields in which the contributions are incremental with respect to existing work (e.g., physics, biology), authors cite their previous research to avoid reproducing the same material in further publications; (iii) due to decreased attention devoted to bibliographical research, sometimes authors cite their own previous work even if less appropriate with respect to other possible references (Schreiber, 2007a); regarding journals: (iv) in niche research fields, self-citations are more common due to the small number of venues the authors can contribute to, and due to the small set of citable documents (Lawani, 1982); finally, (v) to a minor extent, due to the recently explored issue of coercive citations (Wilhite & Fong, 2012).

Regardless the motivations behind self-citations and the debate in the scientific community concerning their extent, most authors agree that a robust indicator should discount self-citations. To date, all methods proposed to discount self-citations assume knowledge on their distribution across papers and rely on their filtering before calculating the impact of a given scholar or journal (Schreiber, 2007a), (Schubert, Glanzel & Thijs, 2006). For example, the ch-index (Kosmulski, 2006) extends the h-index directly ignoring all self-citations. However, for the above-mentioned reasons, we argue that removing all self-citations is not fair, given the fact that they may naturally occur for a number of reasons in all field of science. On the other hand, we state that an unbiased impact measure should deal with self-citations, discounting them when they occur more frequently (perhaps caused by manipulation) or above a certain threshold. For journals evaluation, following this course of action, Thomson Reuters (publisher of the Journal Citation Reports) already suspends indexing a given journal whose number of self-citations grows above a certain threshold. The problem with this approach is to define what is an "acceptable" percentage of self-citations. In fact, common practice is to set an arbitrary upper bound for each category after exploring the data (McVeigh, 2012). The approach presented in this work, however, tries to overcome the need of an exploratory data analysis by considering self-citations as part of the index computation and by proposing a measure that better reflects the impact of both scholars and journals, discounting self-citations.

---

[1] In the case of journals, the provided definition of self-citation must be arranged according to the fact that a journal self-citation is considered as any citation from a paper to another one published in the same venue.

[2] Note that there exist variants of this definition: for example, the one adopted by Thomson Reuters (ISI) Web of Knowledge does not consider as self-citations those provided by co-authors' papers.

The rest of the paper is organized as follows: section 2 presents the ideas underlying our proposal (§ 2.1) and, in addition, all the technical details including the mathematical formulation (§ 2.2), its geometric interpretation (§ 2.3) and some further possible extensions (§ 2.4.)
In section 3 we present the adoption of our impact measure in two real-world cases: (i) the assessment of the impact of the most prolific Computer Science authors (according to the DBLP ranking) as of 2011; and, (ii) the evaluation of the impact of journals belonging to the field "Computer Science Applications" of SCImago. Complementary to the quantitative evaluation, a qualitative analysis is carried out in order to clarify which insights are obtained by means of the adoption of our measure in addition to the original h-index. Finally, in section 4 the paper concludes summarizing our contribution and depicting some potential extensions.

## 2. Methodology

### 2.1. Design Goals

Before a formal description of our index, we illustrate the main ideas behind it, briefly providing a real-life example used to derive some "requirements". Assume an academic committee aims at awarding a grant or an appointment choosing one among a set of potential candidates. Ideally, their evaluation should be as unbiased as possible, so that all candidates are equally considered according to their achievements. Among the criteria, it is a common practice to evaluate publication records and their impact in the scientific community, by means of standard bibliographical indicators, such as the h-index. On the other hand, the h-index suffers from the above-mentioned limitations, including not dealing with self-citations. Hence, a more appropriate choice would be to adopt a similar indicator that preserves the desirable features of h-index while discounting self-citations for an unbiased evaluation.[3] To summarize, the impact measure we want to design should satisfy the following requirements:

**Requirement 1** – *Relatedness to h-index*. Since the h-index is proven to reflect some highly desirable features, it represents an invaluable tool to assess impact in academics, and this is essentially testified by its wide adoption. The proposed index should derive from the h-index, preserving its original characteristics, but including the ability to deal with self-citations.

**Requirement 2** – *Discounting for self-citations*. The presence of self-citations has been explained in different ways, also depending on the research field. Regardless the motivations, it emerges the need for a bibliographical indicator accounting for the presence of self-citations.

**Requirement 3** – *Extensibility to journals*. One of the desirable features of the h-index is its extendibility to assess the impact of journals (Schubert & Glanzel, 2007). The proposed index should extend this ability taking into account the so-called journal self-citations, that is, citations provided by papers published by a given journal to those previously published by the same journal. This could help to put niche publications and journals with broader audience on the same level and discount for the field bias introduced in certain scientific areas.

---

[3] Note that the extension to the evaluation of scientific journals is straightforward.

**Requirement 4** – *Simple computation*. The proposed measure should be a function of the h-index making it possible to compute it without the need of the whole citation data. Several citation database services (e.g., Thomson ISI, SCImago, Scopus, etc.) already provide an overall count of self-citations per author or per journal. The analysis of their distribution is, however, rather difficult, because it involves reviewing each publication. In some cases, it could be impractical and time-consuming to filter out all self-citations for each publication of each considered author.[4] Similar considerations hold for the assessment of the impact of journals belonging to a given field.

In the following, we present a new impact measure that attempts at addressing these requirements.

## 2.2. Discounted h-index

Prior to formally define the new measure, here we introduce the *citation ratio R* between the numbers of genuine citations (i.e., non-self-citations) and overall citations, defined as

$$(1) \quad R = 1 - \frac{SC}{C} = \frac{C-SC}{C},$$

Where $C$ is the overall number of citations, and $SC$ the number of self-citations. It is worth to note that this ratio can be calculated for authors as well as for journals. At this point, we can define the discounted h-index as follows:

**Definition 3.** The discounted h-index (denoted by *dh-index*, or simply *dh*) is defined as the product between the h-index and the square root of the *citation ratio*

$$(2) \quad dh = h \cdot \sqrt{\frac{C-SC}{C}}.$$

For further analysis, we recall here the definition of the indicator known as citation per publication coefficient $C_P$. This coefficient represents the ratio between the overall number of citations and the number of citable documents (of a given author or journal, indifferently), defined as

$$(3) \quad C_P = \frac{C}{CD},$$

Where $CD$ is defined as the number of citable documents and $C$ is again the overall number of citations.
Finally, we hereby derive the adjusted citation per publication coefficient $RC_P$ as the citation per publication coefficient computed with respect to the *citation ratio*, formalized as follows

$$(4) \quad RC_P = R \cdot C_P = \frac{C-SC}{CD}.$$

The derivation of the above-defined index is discussed in the following subsection.

## 2.3. Derivation

An intuitive geometric derivation of our measure is represented in Figure 1. In that figure, $g$ is the curve that shows the number of citations per article,[5] with publications being sorted in decreasing order or citations. $f$ gives the same value but without considering self-citations.

The area under $g$ is $C$, while the area under $f$ is $C - SC$ (whose values are known.) Therefore, the ratio between the areas of $f$ and $g$ is just the *citation ratio R*.

If we denote by $h$ the observed h-index using all the citations $C$, and we define the discounted h-index (namely, *dh*) as the expected h-index without self-citations (whose value is unknown), we can derive the latter assuming that the ratio

---

[4] Some authors observed that it is often sufficient to inspect the self-citations of only a subset of publications for each author, i.e., those with a number of total citations close to the h-index of the given author (Schreiber, 2007b).

[5] Note how this relates to the fact that the square root of the number of citations scales as h (Redner, 2010).

*Emilio Ferrara and Alfonso E. Romero*

between the areas of the squares having lengths *h* and *dh* has the same proportion than their corresponding citation curves. Equation (5) summarizes the assumption

(5) $\frac{dh^2}{h^2} = R.$

The previous equation leads to our definition of the *dh-index* given in equation (2.)

Another interpretation of this index can be given in the framework of Lotkaian systems (Egghe & Rousseau, 2006): as stated in (Egghe, 2010), the variation of the h-index, from $h_{OLD}$ to $h_{NEW}$, considering the removal of a fraction of $k$ sources (the self-citations) is computed as $h_{NEW} = h_{OLD} \cdot \sqrt{1-k}$ that exactly matches our definition of *dh-index*, when the sources removed are precisely the set of self-citations.

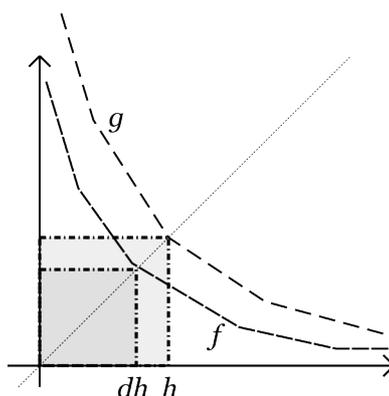

**Figure 1.** Citation curves and their corresponding h-indexes

## 2.4. Possible extensions

The discounted h-index defined in section 2.2, can be seen as a replacement of the original h-index where a scaling factor $\sqrt{R}$ has been introduced as a correction. This correction is a real-value between 0 and 1, computed from the self-citations ratio, and is designed to mitigate their effect over the original index.
The *dh-index* can also be seen as an approximation of the h-index without self-citations (see § 2.3 for that derivation.) If we forget for a moment the theoretical derivation of this index, observing it in this form

(6) $dh = f(R) \cdot h,$

We could then consider a family of general discounted indexes, one for each values of the function $f$ [6] (being $f$ a continuous and derivable real function in [0,1]). Intuitively, not all values of $f$ are reasonable. In fact, $f$ defined as $f: [0,1] \to [0,1]$ should have the two following desirable properties: $f$ should be a monotonically increasing function (we want to penalize the effect of having a greater proportion of self-citations); and, $f(1) = 1$, meaning that there is no discount if no self-citations are observed.

Given those two properties, it can be seen that the (uncorrected) h-index is generalized by the discounted h-index, choosing $f(x) = 1$ (i.e., self-citations are not considered.) Other possible values could be proposed for $f$, falling mainly into these two main categories:[7] (i) concave[8] functions, for instance $f(x) = x^{1/n}, n > 1$ (allowing a certain proportion

---

[6] The "canonic" version of the *dh-index* would be one in which $f$ is the square root of the *citation ratio*.

[7] In fact, an additional comment should be added. There could be functions changing their curvature across the domain (i.e., changing from convex to concave or vice-versa one or more times along [0,1].) We do not find any rationale for the use of this last kind of functions.

[8] Both concavity and convexity can be stated if $f''(x) \geq 0$ or $f''(x) \leq 0$ in the domain (the interval [0,1].)

*Emilio Ferrara and Alfonso E. Romero*

of self-citations without much penalization); and (ii) convex functions, like $f(x) = x^n, n > 1$ (which would highly penalize the occurrence of self-citations).

## 2.5. Discussion

The *dh-index* is not the first attempt to provide a scaled version of the h-index. For example, in (Iglesias & Pecharromán, 2007) the authors propose a coefficient to make h-indexes comparable across different research areas. In addition, this is not the first time that a version of the h-index is scaled using a function of the self-citation rate. In (Brown, 2009) the author presents a derivation of a new measure, called b-index, as a function of the original h-index, the proportion of self-citations and the exponent of a Zipfian relationship. Although a theoretical explanation is presented in the paper, the derivation of that index is not as straightforward, as it needs further computation, and no much evidence is given about its validity. Another argument is that the b-index assumes all citations are distributed according to a Zipf law (Newman, 2005), which does not always hold true, depending on the scientific field, as discussed in recent literature (Radicchi, Fortunato & Vespignani, 2012). Finally, we deem our derivation as more natural and more broadly applicable, as shown is section 3.

## 3. Experimentation

In this section we illustrate two relevant fields of application of our index, which are, respectively, (i) the evaluation of the impact of academic authors (§ 3.1), and (ii) the assessment of the impact of academic journals (§ 3.2.) In addition, we provide an analytical evaluation of the correlation between results and rankings provided by our measure against the original h-index, and the ch-index, i.e., the h-index after filtering out the self-citations, publication per publication (Kosmulski, 2006).

### 3.1. Case 1 – Authors impact evaluation

To prove that the *dh-index* represents a valuable indicator better suited to measure the impact of authors, in the following we provide as a use-case the analysis of the impact of the most prolific Computer Science authors. We provide this as an argument to show that the *dh-index* is a correct and unbiased approximation of the ch-index.
To carry out this experiment we consulted the "most prolific" Computer Scientists list provided by DBLP,[9] as of 2011. Other work carried out similar analysis for non-prominent scientists in fields such as physics (Schreiber, 2007b). For each author appearing in the list, we manually retrieved her/his list of publications, matching her/his profile in Scopus. Together with this list, we obtained the total number of citations, and the number of self-citations to all her/his papers, the h-index and the ch-index, i.e., the indicators of the impact of each author with and without considering the self-citations. After retrieving these data, we were able to compute the *citation ratio* and the corresponding *dh-index* for each of considered authors.
Results related to the top 25 authors are listed in Table 1. In detail, we report: the total number of citable documents (CD) – and the relative ranking; the number of citations (C), self-citations (SC) and the citations per publication coefficient $C_P$; the h-index – and the relative ranking –, together with the ch-index; finally, the *citation ratio R*, the adjusted citation per publication coefficient $RC_P$ and the *dh-index* – together with the relative ranking.

For the sake of simplicity, we discuss the results splitting our analysis in two parts. First, we consider what emerges from the evaluation just reviewing the standard indicators: number of publications, citations per publications coefficient, h-index and ch-index. In the second part of our analysis, we discuss our contribution.

From the analysis of Table 1 it emerges that single criteria (e.g., number of publications or number of citations) are not reliable indicators of the overall impact of each author. It is clearly shown that the ranking according to the total number of publications strongly differ from any other ranking based on multiple criteria (e.g., h-index, *dh-index*.) The citation per publication coefficient does not seem very effective too. In addition, it is evident that this indicator is linearly dependent on the number of self-citations, thus its bias is more marked than that of h-index or ch-index.

---

[9] The list is located in the following URL: http://www.informatik.uni-trier.de/~ley/db/indices/a-tree/prolific/

Indeed, the most interesting insights come from the analysis of these two indicators. For all considered authors, there is a drop in the h-index, after filtering out the self-citations, which ranges from negligible values (≈3%) to considerable values (up to ≈23%.)

The average and median drop in h-index removing the self-citations are respectively ≈10% and ≈9%, with a standard deviation $\sigma = 5.7\%$. Another interesting consideration is that the standard deviation of the h-index distribution is $\sigma = 8.11$, smaller than that of the ch-index which is equal to $\sigma = 8.52$.

In the following, we analyze the contribution provided by our indicator. First of all, we study the *citation ratio* of the considered authors. The *citation ratio* distribution ranges from a minimum of ≈64% (which means that more than one third of the overall citations are self-citations) to a maximum of ≈95%. The average and the median *citation ratios* approximately coincide and are equal to ≈82%, with a standard deviation of ≈8%. This means that, in average for each scholar self-citations account for more or less 20% of overall citations. Considering the adjusted citation per publication coefficient, it clearly emerges that self-citations largely affect this indicator: in fact, the average drop with respect to the non-adjusted coefficient is ≈17% (with a standard deviation of ≈8%), ranging from a minimum of ≈5% to a maximum of ≈35%. These results support the concerns raised by the scientific community on the validity and robustness of the h-index (Costas & Bordons, 2007).

Thus, we consider the *dh-index* and the ranking produced accordingly. First of all, we notice that this ranking is slightly different with respect to that produced by the h-index. As reported in similar studies (Schreiber, 2007b), the presence of self-citations alters the ranking and, in some cases, some authors gain or lose some positions once the self-citations are discounted. On the other hand, the ranking produced by *dh-index* is exactly the same of that produced if we would rank authors with respect to the ch-index, i.e., by filtering out the self-citations.

Interestingly, to produce the *dh-index* we only require the overall number of self-citations, while to produce the ch-index we must be aware of the distribution of self-citations among all the publications; although, it is often sufficient to inspect the publications with a number of citations close to the h-index to obtain an estimate of the ch-index. Since the distribution of self-citations might not be available, or its acquisition might be time consuming, we argue that the utility of the *dh-index* is considerable. Also, using the *dh-index* we are allowing for the influence of "natural" self-citations that normally occur in all fields of science. In the quest of avoiding to discuss per-case author, it is worth noting that, even if no big distortions in the ranking emerge, some authors advanced few positions at the expenses of some others.

### 3.2. Case 2 – Journal impact assessment

The second example provided is a use-case on the evaluation of impact of scientific journals. We selected the set of journals belonging to the category "Computer Science Applications" of the SCImago Journal Ranking (SCImago, 2007). This category, as 2011, contains 194 journals with heterogeneous and broad topics (e.g., bioinformatics, computational methods, information systems, bibliometrics, and much more.) Each journal is provided with a series of statistics, summarized in Table 2 for the top 25 journals, such as the total number of citable documents (CD) – and the relative ranking; the number of citations (C), self-citations (SC) and the citations per publication coefficient $C_P$; the h-index – and the relative ranking. We report, in addition, all statistics related to the *citation ratio*, the adjusted citation per publication coefficient and the *dh-index* – and the relative ranking.

The first important difference that emerges with respect of the previous experiment is that there is not any reference to the ch-index. In fact, in this case there is no actual knowledge on the distribution of the citations among papers published by each journal. Indeed, there is no practical way to infer what would be the h-index obtained after filtering out the self-citations to each paper published by any given journal.

In this context, the contribution of the proposed indicator is clear: we can produce a significant prediction of what would be the h-index of each journal if we would filter out the self-citations, by means of our discounted h-index.

The analysis of Table 2 is as follows: single criteria indicators fail to reflect journals impact (note the fluctuations in the ranking proposed by the number of published papers, or the overall citations.) Similar considerations hold in the case of the citation per publications coefficient. Note that the ranking produced by the *dh-index* is mostly in agreement with that produced by the h-index: a small number of journals gain or lose few positions in the ranking, principally due to the tendency to produce self-citations in their given field.

*Emilio Ferrara and Alfonso E. Romero*

Similarly to the assessment in the case of authors, regarding journals there exists a drop in the h-index with respect to the *dh-index* that ranges from a minimum of ≈2.5% to a maximum of ≈22%. Even though, in case of journals the average and median drop are respectively ≈7.3% and ≈6.4% with a standard deviation of $\sigma = 4.7\%$ – lower than that of authors.

Regarding the *citation ratio*, the distribution ranges from a minimum ≈62% (which means that more than one third of overall citations are self-citations) to a maximum of ≈96%, with an average and a median of respectively ≈86% and ≈87%, and a standard deviation $\sigma = 8.4\%$.

This means that, in spite self-citations alter the h-index and in some case also the ranking of journals, the "average journal" belonging to the category "Computer Science Application" does not suffer of significant ranking bias due to self-citations. As for future work, it would be of extreme interest to track this habit across different scientific fields (as briefly discussed in section 4.)

## 4.  Conclusions

In this paper we presented the *dh-index*, a measure based on h-index to assess the impact of academic research. This measure is devised to discount self-citations in the computation of the impact of a given author or journal. The main strength of this indicator is that it does not require any prior knowledge on the distribution of self-citations among publications. We showed that, in absence of any information about self-citation distributions, our discounted h-index represents a trustworthy prediction of the value that the h-index would acquire if we would be able to filter out self-citations from each publication, but without the drawback of just ignoring all those citations.

We assessed the adoption of *dh-index* in two practical use cases, evaluating authors and journals impact, observing that the *dh-index* is a valuable tool to mitigate the effect of self-citations in the impact evaluation.

As for future work, our ongoing research is focused on the analysis of disparate networks of citations, representing different scientific areas, by adopting the *dh-index*. Our purpose is to assess what type of citation habits different scientific fields exhibit, by analyzing authors and journals citations distribution. This would help to define a general landscape capturing better the differences across scientific fields and allowing for cross-disciplinary comparisons and evaluations. Another potentially interesting direction will be to assess the research impact of countries, whether empirical, non-aggregate data will be made available in the future by scholarly databases.

**Acknowledgments**

The authors are grateful to Filippo Menczer for his helpful comments and suggestions.

*Emilio Ferrara and Alfonso E. Romero*

**Tables**

**Table 1.** Top 25 "Most Prolific Computer Science" authors, reporting citable documents (CD), citations (C), self-citations (SC), citations-per-publication ($C_P$), h-index, ch-index, and our indicators (R, $RC_P$ and *dh-index*.) A * identifies an ex-aequo in the ranking.

| Author | CD | Rank (CD) | C | SC | $C_P$ | h | Rank (h) | ch | R | $RC_P$ | dh | Rank (dh) |
|---|---|---|---|---|---|---|---|---|---|---|---|---|
| **Huang, Thomas** | 784 | 7* | 8,956 | 650 | 11.42 | 44 | 1* | 43 | 0.92 | 10.59 | **42.37** | 1 |
| **Wang, Wei** | 1,183 | 4 | 10,723 | 1,558 | 9.06 | 44 | 1* | 41 | 0.85 | 7.74 | **40.68** | 2 |
| **Yu, Philip** | 392 | 20 | 5,348 | 305 | 13.64 | 41 | 3 | 40 | 0.94 | 12.86 | **39.81** | 3 |
| **Poor, H. Vincent** | 784 | 7* | 8,501 | 955 | 10.84 | 39 | 4* | 38 | 0.88 | 9.62 | **36.74** | 4 |
| **Han, Jiawei** | 302 | 24 | 5,985 | 299 | 19.81 | 37 | 6 | 36 | 0.95 | 18.82 | **36.06** | 5 |
| **Li, Ming** | 1,542 | 2 | 11,049 | 1,946 | 7.16 | 39 | 4* | 34 | 0.82 | 5.90 | **35.40** | 6 |
| **Shin, Kang** | 541 | 13 | 4,898 | 248 | 9.05 | 35 | 8 | 34 | 0.94 | 8.59 | **34.10** | 7 |
| **Li, Xin** | 1,568 | 1 | 8,591 | 1,648 | 5.47 | 36 | 7 | 33 | 0.80 | 4.42 | **32.36** | 8 |
| **Seidel, Hans Peter** | 351 | 23 | 4,427 | 603 | 12.61 | 31 | 10* | 29 | 0.86 | 10.89 | **28.81** | 9 |
| **Sangiovanni-Vincentelli, Alberto** | 572 | 12 | 3,239 | 398 | 5.66 | 30 | 12* | 28 | 0.87 | 4.96 | **28.10** | 10 |
| **Chang, ChinChen** | 704 | 10 | 4,300 | 794 | 6.10 | 31 | 10* | 28 | 0.81 | 4.98 | **27.99** | 11 |
| **Bertino, Elisa** | 372 | 21 | 3,296 | 353 | 8.86 | 29 | 14 | 28 | 0.89 | 7.91 | **27.40** | 12 |
| **Pedrycz, Witold** | 752 | 9 | 4,643 | 1,262 | 6.17 | 32 | 9 | 28 | 0.72 | 4.49 | **27.31** | 13 |
| **Wang, Jun** | 1,038 | 5 | 4,892 | 971 | 4.71 | 30 | 12* | 27 | 0.80 | 3.77 | **26.86** | 14 |
| **Gao, Wen** | 887 | 6 | 3,901 | 509 | 4.39 | 26 | 15 | 24 | 0.87 | 3.82 | **24.25** | 15 |
| **Kandemir, Mahmut Taylan** | 424 | 18 | 2,350 | 275 | 5.54 | 23 | 19 | 21 | 0.88 | 4.89 | **21.61** | 16 |
| **Abraham, Ajith** | 368 | 22 | 2,239 | 426 | 6.08 | 24 | 16 | 21 | 0.81 | 4.92 | **21.60** | 17 |
| **Zhang, Yan** | 607 | 11 | 2,237 | 338 | 3.68 | 23 | 17* | 21 | 0.84 | 3.12 | **21.19** | 18 |
| **Wu, Jie** | 463 | 16 | 2,189 | 387 | 4.72 | 23 | 17* | 21 | 0.82 | 3.89 | **20.87** | 19 |
| **Liu, Yang** | 1,213 | 3 | 3,712 | 788 | 3.06 | 22 | 21* | 20 | 0.78 | 2.41 | **19.53** | 20 |
| **Hancock, Edwin R.** | 462 | 17 | 2,337 | 677 | 5.05 | 23 | 17* | 19 | 0.71 | 3.59 | **19.38** | 21 |
| **Reddy, Sudhakar M.** | 529 | 15 | 2,120 | 616 | 4.00 | 22 | 21* | 17 | 0.70 | 2.84 | **18.53** | 22 |
| **Liu, Wei** | 532 | 14 | 1,825 | 493 | 3.43 | 19 | 23 | 15 | 0.73 | 2.50 | **16.23** | 23 |
| **Rozenberg, Grzegorz** | 288 | 25 | 864 | 183 | 3.00 | 17 | 24 | 15 | 0.78 | 2.36 | **15.09** | 24 |
| **Piattini, Mario** | 394 | 19 | 1,182 | 423 | 3.00 | 16 | 25 | 13 | 0.64 | 1.92 | **12.82** | 25 |

*Emilio Ferrara and Alfonso E. Romero*

**Table 2.** Top 25 "Computer Science Applications" journals, reporting citable documents (CD), citations (C), self-citations (SC), citations-per-publication ($C_P$), h-index, and our indicators (R, $RC_P$ and *dh-index*.) Journal names are abbreviated using the standard ISI nomenclature. A * identifies an ex-aequo in the ranking.

| Journal | CD | Rank (CD) | C | SC | $C_P$ | h | Rank (h) | R | $RC_P$ | dh | Rank (dh) |
|---|---|---|---|---|---|---|---|---|---|---|---|
| **Bioinformatics** | 2,104 | 3 | 6,510 | 567 | 2.93 | 176 | 1 | 0.91 | 2.67 | **168.16** | **1** |
| **IEEE T Med Imaging** | 503 | 20 | 1,229 | 107 | 2.24 | 116 | 2* | 0.91 | 2.04 | **110.84** | **2** |
| **J Comput Phys** | 1,387 | 4 | 2,595 | 622 | 1.62 | 116 | 2* | 0.76 | 1.23 | **101.15** | **3** |
| **Comput Method Appl M** | 868 | 8 | 1,626 | 300 | 1.74 | 92 | 4 | 0.81 | 1.41 | **83.08** | **4** |
| **J Chem Inf Model** | 697 | 11 | 1,839 | 286 | 2.65 | 90 | 5 | 0.84 | 2.23 | **82.71** | **5** |
| **BMC Bioinformatics** | 2,192 | 2 | 3,881 | 335 | 1.52 | 78 | 6* | 0.91 | 1.38 | **74.56** | **6** |
| **Comput Phys Commun** | 721 | 10 | 1,151 | 109 | 1.68 | 78 | 6* | 0.90 | 1.52 | **74.22** | **7** |
| **IEEE T Comput Aid D** | 670 | 13 | 568 | 73 | 0.63 | 70 | 8 | 0.87 | 0.54 | **65.35** | **8** |
| **Comput Chem Eng** | 610 | 14 | 1,089 | 160 | 1.7 | 68 | 9 | 0.85 | 1.45 | **62.81** | **9** |
| **Med Image Anal** | 190 | 59 | 582 | 44 | 2.71 | 60 | 11 | 0.92 | 2.50 | **57.69** | **10** |
| **Comput Struct** | 461 | 24 | 667 | 61 | 1.3 | 54 | 14 | 0.90 | 1.18 | **51.47** | **11** |
| **Int J Numer Meth Fl** | 675 | 12 | 587 | 76 | 0.74 | 55 | 13 | 0.87 | 0.64 | **51.31** | **12** |
| **Inform Process Manag** | 222 | 50 | 277 | 12 | 0.91 | 51 | 16 | 0.95 | 0.87 | **49.88** | **13** |
| **IET Control Theory Appl** | 489 | 21 | 411 | 51 | 0.66 | 53 | 15 | 0.87 | 0.57 | **49.60** | **14** |
| **Inform Sciences** | 1,008 | 5 | 2,595 | 921 | 2.32 | 61 | 10 | 0.64 | 1.49 | **48.99** | **15** |
| **IEEE T Syst Man Cy C** | 175 | 62 | 328 | 17 | 1.39 | 49 | 19 | 0.94 | 1.31 | **47.71** | **16** |
| **J Chem Theory Comput** | 877 | 7 | 2,880 | 380 | 2.85 | 50 | 17 | 0.86 | 2.47 | **46.59** | **17** |
| **Expert Syst Appl** | 2,773 | 1 | 6,064 | 2290 | 2.08 | 57 | 12 | 0.62 | 1.29 | **44.97** | **18** |
| **Med Biol Eng Comput** | 383 | 30 | 446 | 71 | 1.14 | 48 | 20 | 0.84 | 0.95 | **44.01** | **19** |
| **Cyberpsychol Behav Soc Netw** | 336 | 38 | 510 | 53 | 1.32 | 46 | 22* | 0.89 | 1.18 | **43.54** | **20** |
| **IEEE T Inf Technol B** | 373 | 31 | 440 | 40 | 1.04 | 45 | 26 | 0.90 | 0.94 | **42.91** | **21** |
| **Struct Multidiscip O** | 408 | 28 | 395 | 82 | 0.98 | 47 | 21 | 0.79 | 0.77 | **41.84** | **22** |
| **Comput Ind** | 205 | 54 | 338 | 59 | 1.29 | 46 | 22* | 0.82 | 1.06 | **41.79** | **23** |
| **J Syst Software** | 536 | 19 | 583 | 48 | 0.85 | 43 | 27 | 0.91 | 0.78 | **41.19** | **24** |
| **J Parallel Distr Com** | 302 | 44 | 289 | 16 | 0.75 | 41 | 29 | 0.94 | 0.70 | **39.85** | **25** |